\documentclass[pdflatex,sn-nature]{sn-jnl}

\usepackage{graphicx}%
\usepackage{multirow}%
\usepackage{amsmath,amssymb,amsfonts}%
\usepackage{amsthm}%
\usepackage{mathrsfs}%
\usepackage[title]{appendix}%
\usepackage{xcolor}%
\usepackage{textcomp}%
\usepackage{manyfoot}%
\usepackage{booktabs}%
\usepackage{algorithm}%
\usepackage{algorithmicx}%
\usepackage{algpseudocode}%
\usepackage{listings}%
\usepackage{array}
\usepackage{credits}
\usepackage{comment}

\theoremstyle{thmstyleone}%

\theoremstyle{thmstyletwo}%

\theoremstyle{thmstylethree}%

\definecolor{FVAcolor}{RGB}{0,102,204}
\definecolor{CPMcolor}{RGB}{204,0,102}
\definecolor{ERFcolor}{RGB}{204,0,0}

\credit{ERF}{1, 1, 1, 0, 1, 1, 0, 0, 1, 0, 1, 1, 1, 1}
\credit{CPM}{1, 0, 0, 1, 0, 0, 1, 1, 0, 1, 1, 0, 0, 1}
\credit{FVA}{1, 0, 1, 1, 1, 1, 1, 1, 1, 1, 1, 0, 0, 1}
\credit{SR}{ 0, 0, 0, 1, 0, 0, 1, 1, 0, 1, 1, 0, 0, 1}

\input{Galois_connections}
\input{maxplus_minplus}

\newcommand{\Rbar}{\overline{\mathbb{R}}}
\newcommand{\mplus}{\mathbin{\underaccent{\centerdot}{\oplus}}}
\newcommand{\mtimes}{\mathbin{\underaccent{\centerdot}{\otimes}}}
\newcommand{\nplus}{\mathbin{\overset{\centerdot}{\oplus}}}
\newcommand{\ntimes}{\mathbin{\overset{\centerdot}{\otimes}}}
\newcommand{\mstar}{\underaccent{\centerdot}{\circledast}}
\newcommand{\msum}[1]{\sum\nolimits_{\bullet #1}}
\newcommand{\mpow}[1]{^{\underaccent{\centerdot}{\otimes} #1}}
\newcommand{\tr}{^{\mathsf{T}}}

\begin{document}

\title[Every route counts, but some count more: a max-plus view of the \textit{C.~elegans} brain]{Every route counts, but some count more: a max-plus view of the \textit{C.~elegans} brain}

\author*[1,2]{\fnm{Emma} \sur{Reyner-Fuentes}}\email{emma.reyner@uc3m.es}

\author[3]{\fnm{Sebastian} \sur{Rudolph}}\email{sebastian.rudolph@tu-dresden.de}

\author[1]{\fnm{Carmen} \sur{Peláez-Moreno}}\email{carmen@tsc.uc3m.es}

\author[2]{\fnm{Francisco J.} \sur{Valverde-Albacete}}\email{francisco.valverde@urjc.es}

\affil[1]{\orgdiv{Department of Signal Theory and Communications}, \orgname{University Carlos III of Madrid (UC3M}), \orgaddress{\city{Legan\'es}, \state{Madrid}, \country{Spain}}}

\affil[2]{\orgdiv{Department of Signal Theory and Communications and Telematic Systems and Computation}, \orgname{University Rey Juan Carlos}, \orgaddress{\city{Fuenlabrada}, \state{Madrid}, \country{Spain}}}

\affil[3]{\orgdiv{Computational Logic Group, Institute for Artificial Intelligence}, \orgname{Technische Universität Dresden}, \orgaddress{\city{Dresden}, \state{Saxony}, \country{Germany}}}

\abstract{Brains do not simply add up every signal they receive: they select which routes dominate. Yet connectomes are usually analyzed with tools that sum over all paths and place neural states in Euclidean spaces, blurring the feedback circuits that may sustain internal states. Here we treat neural signaling as routing instead of diffusion and describe it with max-plus algebra, in which signals accumulate along a path and the strongest route prevails. We propose a model in which the functional connectome of \textit{Caenorhabditis elegans} emerges from the time-varying combination of electrical, synaptic, monoaminergic, neuropeptidergic and hormonal layers, and analyze five networks within this framework. Each collapses onto a handful of dominant feedback loops, which recover well-characterized circuits and point to new candidates for experimental study. Unlike linear spectra, these structures keep the biological meaning of the data, and their order-based description suggests that brain states live in lattices rather than Euclidean spaces.}


\maketitle

\section{Introduction}\label{sec1}

Understanding how nervous systems actually function is a central objective of neuroscience. A systems-level explanation requires relating the properties of individual neurons to the organization that emerges through their interactions. These interactions allow signals to propagate through the network and return through recurrent circuits, contributing to the ongoing evolution of neuronal activity. Determining how this organization arises from different mechanisms of communication, and how it changes over time, is an essential part of the effort to reverse engineer a nervous system~\cite{haspel2023}. This perspective places the relationship between connectivity and internal dynamics at the center of the analysis.

The nematode \emph{Caenorhabditis elegans} provides a particularly tractable system in which to investigate this relationship. Its adult hermaphrodite contains 302 neurons, most of which can be individually identified across animals and experimental datasets. Reconstructions of its anatomical connectome describe the chemical synapses and electrical gap junctions linking these cells~\cite{white1986,cook2019}. Complementary maps characterize potential communication through monoamines and neuropeptides, while studies of hormonal signaling reveal additional mechanisms through which physiological conditions influence nervous-system function~\cite{bentley2016,ripollsanchez2023,istiban2024}. Together, these resources support the study of several forms of neuronal communication within a common cellular reference system, allowing network-level properties to be traced back to identified neurons and their interactions.

These communication mechanisms, however, differ in both signal-transmission times and the duration of their effects. Gap junctions permit direct ionic current between coupled cells, supporting electrical transmission with minimal delay. Fast chemical synaptic transmission involves neurotransmitter release and postsynaptic receptor activation, typically operating on a millisecond timescale. Monoaminergic signaling can modulate neuronal activity over seconds to minutes, while neuropeptidergic signaling spans rapid responses and sustained effects lasting minutes or longer~\cite{bentley2016,flavell2013,randi2023}. Hormonal pathways further link neuronal operation to physiological processes whose temporal properties depend on the particular signaling system~\cite{istiban2024}. These temporal properties determine when each mechanism contributes to neuronal interaction and how long its effects persist.

For a model of neuronal interaction, both the onset and persistence of these effects matter. Rapid transmission can modify the immediate propagation of activity, while sustained modulation can change the contribution of particular connections over longer intervals. The resulting network therefore depends on which communication mechanisms are engaged and how their effects develop and decay. This temporal organization motivates a description in which the effective contribution of each connectivity layer varies over time.

These combined influences can be described through a \emph{functional connectome}, understood here as the effective network of directed neuronal interactions under particular conditions. An experimental view of this network is provided by the signal-propagation atlas of Randi et al., which combines stimulation of individual neurons with whole-brain calcium imaging to measure responses in other identified cells~\cite{randi2023}. The atlas captures interactions mediated by the communication routes engaged under the experimental conditions, including polysynaptic and extrasynaptic effects. Its network architecture differs from that of the anatomical connectome, illustrating how structural wiring and measured signal propagation provide complementary descriptions of neuronal organization~\cite{dvali2025}. Within a dynamical model, the functional connectome can consequently be viewed as a time-dependent operator whose organization reflects the contributions of the underlying signaling mechanisms.

The relevance of this temporal organization becomes apparent when considering internal states. In \emph{C.~elegans}, for example, persistent roaming and dwelling states are regulated by serotonergic and peptidergic signaling, linking changes in neuronal communication to sustained modes of operation~\cite{flavell2013}. Within and across behavioral states, population activity evolves through coordinated trajectories, usually described in low-dimensional Euclidean embeddings, and individual neurons can change how they encode behavior~\cite{kato2015,atanas2023}. Models of whole-brain activity further describe effective interactions through combinations of dynamical operators whose contributions vary with context~\cite{yezerets2025}. These observations motivate the hypothesis that the persistence and reconfiguration of internal states may depend in part on the recurrent organization of the functional connectome. Examining that organization provides a way to identify the neuronal circuits and communication mechanisms that could support such regimes.

A useful mathematical description of this organization should make it possible to follow interactions from individual connections to complete neuronal circuits. Existing analyses of the \emph{C.~elegans} connectome have identified communication pathways, related neuronal position to network controllability, and used spectral representations to describe large-scale connectivity~\cite{bacik2016,yan2017,petrovic2019}. For the model considered here, a particularly relevant feature is feedback: directed routes through which an influence can return to a neuron or group of neurons. The weights of the connections along these routes determine which loops are most strongly represented according to the measurements encoded in the network. Their location and connections to the surrounding neurons provide a way to examine how local recurrent structures are embedded within a whole-system architecture. Changes in the available connections or their weights can then alter which recurrent routes dominate under a specified rule for combining interactions.

Most of these analyses, however, rest on assumptions that limit what they can say about feedback. Community detection and rich-club analyses usually work on undirected, often unweighted, representations and frequently merge chemical and electrical contacts into a single graph~\cite{emmons2024,onuchin2023,towlson2013}; they describe segregation and integration, but not which modules are recurrent or how they are ordered along the direction of propagation. Linear spectral methods and centrality measures keep direction and weight, but add the contributions of all walks, so their leading mode is dominated by densely connected hubs and their remaining modes mix many neurons with signed or complex coordinates~\cite{seguin2023,varshney2011}. What is missing is a description that takes direction and weight into account, keeps each communication mechanism separate, and returns the specific closed routes that dominate each network.

A further assumption concerns the space in which neural states are described. Whole-brain activity is usually represented as points of a Euclidean space, in which activity patterns combine by weighted superposition, their similarity is measured by distances, and structure is sought in linear subspaces or low-dimensional manifolds~\cite{kato2015,jolliffe2016}. This representation carries an implicit model of how signals propagate. In the taxonomy of network communication models, connectome signalling ranges from diffusion, in which influence spreads along all available walks and accumulates additively, to routing, in which signals follow selected optimal paths~\cite{seguin2023}. A Euclidean state space is natural only at the diffusive end of this spectrum. Selective forms of polysynaptic signalling, such as thresholding, competition between inputs, and the gating of one pathway by another, choose among alternative routes rather than add them, so the states they produce are not closed under superposition. Moreover, many questions about internal states are questions of order rather than distance: which neurons are engaged together, which activity patterns contain others, and which patterns are indistinguishable to downstream neurons. A Euclidean space provides a metric but no intrinsic hierarchy between qualitatively distinct states. These properties call for a state space organized by order and selection, a lattice in which patterns are compared by inclusion and any two patterns have a well-defined least upper bound and greatest lower bound.

Max-plus algebra offers a direct way to formulate this perspective. It is a tropical algebra in which maximization replaces ordinary addition and ordinary addition replaces multiplication~\cite{gondran2008,gaubert1997}. Applied to a weighted network, these operations accumulate weights along successive connections and select the largest accumulated weight among alternative routes. Max-plus algebra is therefore the algebra of the routing end of the communication spectrum. Its order dual, min-plus algebra, replaces maximization by minimization and is the algebra of shortest-path routing; the two are exchanged by a change of sign, so that selecting the strongest route on weights is equivalent to selecting the shortest route on costs. Together they make the spaces of states ordered by construction: the maximum of two activity patterns is their least upper bound and their minimum is their greatest lower bound, so these spaces form lattices rather than Euclidean vector spaces. This makes dominant paths and recurrent loops explicit objects of the analysis. Its value for the present problem lies in linking a network-level description to identifiable feedback circuits and in providing an algebraic rule through which changes in connectivity can change the dominant recurrent organization.

This path-based interpretation also gives spectral analysis a concrete biological meaning. Max-plus spectral quantities identify recurrent loops with the greatest average edge weight and describe their relation to the neurons connected to them. The interpretation of that weight follows the biological quantity represented in each dataset, such as anatomical contact extent, supported ligand--receptor interactions, or measured response amplitude. The resulting description connects compact recurrent circuits with the larger neuronal domains that can influence them or be reached from them. The generalized framework also accommodates networks containing several recurrent regions, preserving their distinct organization and directed relationships~\cite{valverde2016}. These properties make max-plus suitable for examining a model in which signaling mechanisms acting over different timescales alter the relative prominence of recurrent routes.

Tropical algebra has also been explored as a tool for modelling neuronal networks. In a preprint, Patel~\cite{patel2021} proposed a threshold model of timing dynamics, in which each neuron updates its state once a fraction of its inputs has arrived, and used its max-plus formulation as a first attempt to extract backbone structures from the \emph{C.~elegans} neuronal network. That approach simulates activity on a single wiring diagram; here, we instead analyze the spectral structure of the connectivity itself, separately for each communication layer.

We therefore propose a model in which the time-dependent functional connectome emerges from the max-plus composition of electrical, chemical-synaptic, monoaminergic, neuropeptidergic, and hormonal connectivity layers, each engaged to a different extent over time. Within this model, the activation of a signaling layer can merge recurrent neuronal groups or change which feedback circuit dominates within them, providing a candidate mechanism for the reconfiguration of internal states.
 
The available datasets measure different biological quantities, from anatomical contact extent to ligand--receptor compatibility and evoked calcium responses, and cannot yet be combined into this composite operator. We therefore analyze the five available networks (gap junctions, chemical synapses, monoaminergic and neuropeptidergic signaling, and the experimentally measured signal-propagation atlas) separately within the same max-plus framework. For each one we identify its recurrent domains, its dominant feedback circuits, the upstream and downstream neuronal populations attached to them, and we relate these structures to known neuronal function.
 
This paper focuses on the biological application and presents the method at the level needed to interpret the results. The underlying theory, in which the nervous system is treated as a dynamical system over a dual pair of max-plus and min-plus semifields whose state space is organized as a hierarchy of lattices rather than a Euclidean space, is developed in a companion paper~\cite{companion}.

\section{Results}\label{sec2}

In this section, we first present a simplified version of the theoretical model, in which the functional connectome is a time-dependent composition of neuronal communication layers, and describe its implications for recurrent organization. This version is sufficient to interpret and apply the method and to show what it adds to the analysis of neuronal connectivity; the full mathematical development of the model is given in a companion paper~\cite{companion}. We then move to empirical measurements and examine the electrical, chemical-synaptic, monoaminergic, neuropeptidergic, and experimentally measured functional networks individually, identifying their dominant recurrent circuits and their relationships with the surrounding neuronal populations. Finally, we compare these structures with conventional connectivity analyses.

\subsection{The model}
\label{sec:model}

We consider the internal organization of the nervous system through a state-transition model in which neuronal interactions determine how the state evolves over discrete time $t\in\mathbb{N}$:
\begin{equation}
x(t+1) = A(t) \mtimes x(t), \qquad x_i(t+1) = \max_{j}\bigl(A_{ij}(t) + x_j(t)\bigr), \label{eq:dynamics}
\end{equation}
where $x(t)$ is the modeled neuronal state and $A(t)$ the functional connectome at time $t$, with $A_{ij}(t)$ the weight of the influence from neuron $j$ to neuron $i$. Here $\mtimes$ denotes the max-plus matrix product, in which ordinary sums are replaced by maximization and ordinary products by addition; operators with a dot below belong to the completed max-plus semifield $\cmaxplus$, whose entries take values in $\Rbar=\mathbb{R}\cup\{\pm\infty\}$ (Section~\ref{sec:maxplus}). For each neuron, the update therefore selects the largest contribution obtained by adding an incoming connection weight to the state of its source neuron. Equation~\eqref{eq:dynamics} describes interactions among the 302 neurons and focuses on the internal state of the nervous system, without environmental inputs or feedback from behavior; these can be added to describe the complete sensorimotor loop~\cite{companion}.
 
The functional connectome is supported by several superposed mechanisms of communication. Let $A^{(G)}$, $A^{(S)}$, $A^{(M)}$, $A^{(N)}$ and $A^{(H)}$ denote the contributions of gap junctions, chemical synapses, monoaminergic, neuropeptidergic, and hormonal signaling, respectively. We propose that the functional operator is their gated composition
\begin{equation}
A(t) = \msum{\ell}\, \alpha_\ell(t) \mtimes A^{(\ell)}, \qquad A_{ij}(t) = \max_{\ell}\bigl(\alpha_\ell(t) + A^{(\ell)}_{ij}\bigr), \qquad \ell \in \{G,S,M,N,H,\dots\}, \label{eq:composition}
\end{equation}
where the scalar gate $\alpha_\ell(t)\in\Rbar$ expresses how strongly layer $\ell$ is engaged at time $t$% ($\alpha_\ell=\bot$: off; $\alpha_\ell=e=0$: fully on)
. For each ordered pair of neurons, the composite retains the largest gated contribution among the layers. The formulation assumes that all layers are expressed on a common scale and remains open to further communication mechanisms. Over the operational timescale considered we treat electrical coupling as a stable scaffold ($\alpha_G\equiv e$), whereas the remaining gates switch on and decay with the kinetics of each signaling mechanism. The same anatomical and molecular resources can therefore contribute differently to the functional connectome at different times.
 
This composition has three consequences for the organization of the modeled nervous system (formal statements in~\cite{companion}).
\begin{enumerate}
\item \textit{Engaging a layer can merge recurrent groups, but never split them.} The graph of $A(t)$ is the union of the graphs of the engaged layers. Neurons that can reach each other through electrical coupling therefore remain mutually reachable, and connections contributed by other layers can close new routes between groups. The partition of neurons into recurrent components is always equal to, or coarser than, that of $A^{(G)}$.
\item \textit{Engaging a layer can change which circuit dominates.} A layer can strengthen an existing connection or add a new one, changing the weights accumulated around closed routes. Within a recurrent component, the closed routes with the greatest average connection weight are its \emph{critical cycles}; a change in the gates can change them even when the component keeps the same neurons. Because adding the same constant to every gate shifts every cycle mean by that constant, the critical cycles depend only on the \emph{relative} engagement of the layers, $\alpha_\ell(t)-\alpha_m(t)$.
\item \textit{Signaling kinetics set the time course of this reorganization.} Fast synaptic changes can alter which cycles dominate over short intervals, whereas sustained monoaminergic, neuropeptidergic or hormonal signaling can shift their relative weights for minutes or longer. As gates rise and decay, different cycles can become critical in turn.
\end{enumerate}
 
The spectral description extends these consequences from the critical circuits to the rest of the network. Max-plus eigenvectors characterize the weighted pathways connecting a critical circuit with other neurons: left eigenvectors describe its \emph{upstream} domain, the neurons from which signals can reach the circuit, and right eigenvectors its \emph{downstream} domain, the neurons the circuit can reach. Changes in the critical circuits are therefore accompanied by changes in the neuronal domains organized around them. Within the model, sustained configurations of these interactions could contribute to the maintenance of internal states, and their reconfiguration to transitions between states such as roaming and dwelling~\cite{flavell2013}. % TODO key [7]
 
The functional signal-propagation atlas plays a distinct role in this model: it is not a further layer, but an experimental observation of the composite operator under particular conditions, including polysynaptic effects.
 
Overall, the model links the kinetics of different communication mechanisms to the recurrent organization of the functional connectome, from its critical circuits to their upstream and downstream domains. In the next subsections, we examine this organization in the available datasets through independent max-plus analyses, which preserve the biological meaning of each dataset while placing their recurrent architectures in a common mathematical framework. Finally, we compare the max-plus description with conventional connectivity analyses of the same networks, including linear spectral methods, community detection and rich-club analysis, to show what it adds to their view of the connectome.

\subsection{Empirical analysis}
\label{sec:empirical_analysis}

The composition proposed in Section~\ref{sec:model} requires the contributions of the different communication layers to be quantitatively comparable. To examine the available evidence, we assembled five complementary network descriptions of the \emph{C.~elegans} nervous system. Chemical-synapse and gap-junction connectivity were obtained from the anatomical reconstruction of Cook et al.~\cite{cook2019}. The monoaminergic network describes potential signaling relationships based on neuronal transmitter identities and compatible receptor expression~\cite{bentley2016,wang2024}, while the neuropeptidergic network combines peptide and receptor expression with experimentally characterized ligand--receptor relationships and spatial constraints~\cite{ripollsanchez2023}. The functional signal-propagation network was obtained from the optogenetic stimulation and calcium-imaging experiments of Randi et al.~\cite{randi2023}. Dataset versions, coverage, and preparation are described in Section~\ref{sec:connectivity_datasets}.

These networks were reconstructed using different experimental approaches, and their weights represent different biological quantities. Anatomical contact extent, the number of supported transmitter--receptor relationships, and calcium-response amplitude each have their own units and interpretation. Their direct combination would therefore compare numerical values with different biological meanings. The datasets also differ in experimental conditions and neuronal coverage, providing separate descriptions of connectivity instead of the simultaneous measurements of the contributions required to reconstruct $A(t)$.

Each network can nevertheless be characterized independently within the proposed max-plus framework. This allows us to demonstrate its usefulness for understanding the system-level organization of neuronal connectivity while preserving the biological meaning of each dataset. The empirical analysis therefore examines the five available networks individually; hormonal signaling remains part of the theoretical model only, because no neuron-to-neuron hormonal map is yet available.

For each network, we examine the following complementary structures. Appendix~\ref{app:maxplus_guide} provides a quick reading guide to their mathematical meaning and biological interpretation.

\begin{itemize}
    \item \textbf{Strongly connected components (SCCs)} identify maximal groups of neurons that can be mutually reached through directed paths. Components containing cycles delimit the regions in which recurrent interaction is structurally possible.

    \item \textbf{The condensation graph} represents each SCC as a single node, revealing how recurrent regions and peripheral neurons are connected. It provides a compact description of the upstream and downstream organization of the network.

    \item \textbf{Critical cycles and critical classes} identify the closed neuronal routes with the greatest mean connection weight within a recurrent component. Their organization reveals the specific circuits that dominate under the max-plus interpretation.

    \item \textbf{The eigenvalue $\lambda$} gives the maximum mean weight of these recurrent routes. Its biological meaning follows the quantity encoded by the edge weights in the corresponding dataset.

    \item \textbf{Right and left eigenvectors} extend the description of critical circuits to their surrounding neuronal domains. Right eigenvectors characterize the downstream organization reachable from a critical circuit, while left eigenvectors characterize the upstream organization from which it can be reached. Their coordinates describe these relationships in terms of weighted pathways.
\end{itemize}

Together, these structures reveal where recurrence is possible, which weighted circuits dominate within those regions, and how those circuits are connected to the wider neuronal population. Throughout the analysis, an entry $A_{ij}$ represents a connection from neuron $j$ to neuron $i$, establishing the direction used to interpret upstream and downstream relationships.

\subsubsection{Gap-junction connectome}
\label{sec:maxplus_gapjunction}
\begin{figure}
    \centering
    \includegraphics[width=0.5\linewidth]{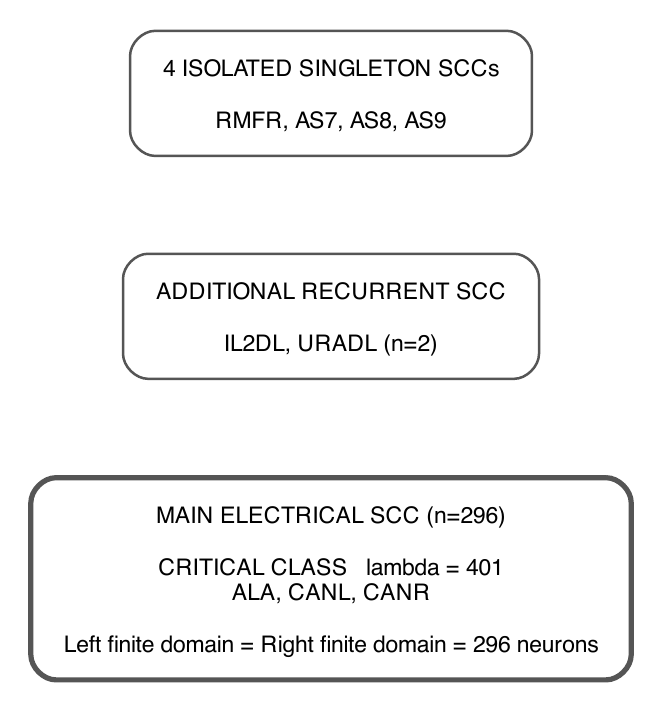}
    \caption[Schematic organization of the gap-junction connectome]{Schematic organization of the gap-junction connectome}
    \label{fig:gap_junctions_architecture}
\end{figure}

The electrical connectome provides the first view of recurrent organization across the individual connectivity layers. Gap junctions are bidirectional and, therefore, represented as symmetric connections in the Cook et al. matrix~\cite{cook2019}. Consequently, the right and left eigensystems are identical and strongly connected components coincide with ordinary connected components. There is therefore no distinct upstream--downstream organization to recover in this layer, and separate right/left representations or a condensation graph would be redundant. Nevertheless, a schematic representation of the connectome is presented in Figure~\ref{fig:gap_junctions_architecture} in order to provide a global overview of its organization. 

At the global level, electrical connectivity is dominated by a single component containing 296 of the 302 neurons. A second component contains only IL2DL and URADL, while RMFR, AS7, AS8, and AS9 occur as four isolated singleton components. Thus, almost the entire nervous system belongs to one mutually reachable electrical domain, in sharp contrast with the directed hierarchy observed for chemical synapses. 

Two finite spectral modes are obtained. The dominant mode has eigenvalue $\lambda=401$ and spans the 296-neuron main component. Despite the large size of this connected domain, its critical class contains only three neurons: ALA, CANL, and CANR. The corresponding critical graph consists of four electrical edges,
\[
\mathrm{ALA}\leftrightarrow\mathrm{CANL},
\qquad
\mathrm{ALA}\leftrightarrow\mathrm{CANR},
\]
each with weight 401. The critical class therefore contains two equivalent length-two cycles, ALA--CANL--ALA and ALA--CANR--ALA, both with cycle mean 401. 
The magnitude of this eigenvalue has to be interpreted in the units of the anatomical reconstruction. In the Cook et al. gap junction matrices, edge weights correspond to the total number of serial electron-microscopy sections over which connectivity is observed, incorporating both the number and anatomical extent of the contacts~\cite{cook2019}. An eigenvalue of 401 therefore identifies an exceptionally extensive recurrent electrical contact in the reconstructed anatomy; it is not a measure of electrical current, neuronal activity, or physiological coupling strength.

The second spectral mode is much smaller: it has eigenvalue $\lambda=1$ and corresponds to the isolated two-neuron component formed by IL2DL and URADL. The two neurons are connected reciprocally with weight 1, so the complete component is also its critical class:
\[
\mathrm{IL2DL}\xrightarrow{1}\mathrm{URADL}
\xrightarrow{1}\mathrm{IL2DL}.
\]
IL2DL belongs to the IL2 sensory class, whose neurons play a well-established role in the dauer-specific dispersal behaviour known as nictation~\cite{lee2012nictation}. The specific physiological role of the IL2DL--URADL electrical coupling has not, however, been established, and the present result is therefore interpreted primarily as a structurally independent electrical recurrent component.

Overall, the gap-junction network combines broad electrical integration with a small number of sharply defined recurrent structures. Almost the entire nervous system belongs to a single SCC, within which the dominant spectral class is supported by the reciprocal ALA--CANL/R couplings. A second spectral class is associated with the isolated IL2DL--URADL component, showing that recurrent electrical organization is not restricted to the large connected domain. The resulting picture is therefore one of a largely unified electrical network whose spectral organization is concentrated in a few highly specific neuronal pairings.

\subsubsection{Synaptic connectome}
\label{sec:maxplus_synaptic}

The chemical-synapse connectome provides a markedly different view of neuronal organization from the electrical network examined above. Its directed synaptic architecture provides a direct anatomical view of how recurrent neuronal organization emerges from the wiring of the nervous system.

\begin{figure}
    \centering
    \includegraphics[
        width=\linewidth,
    ]{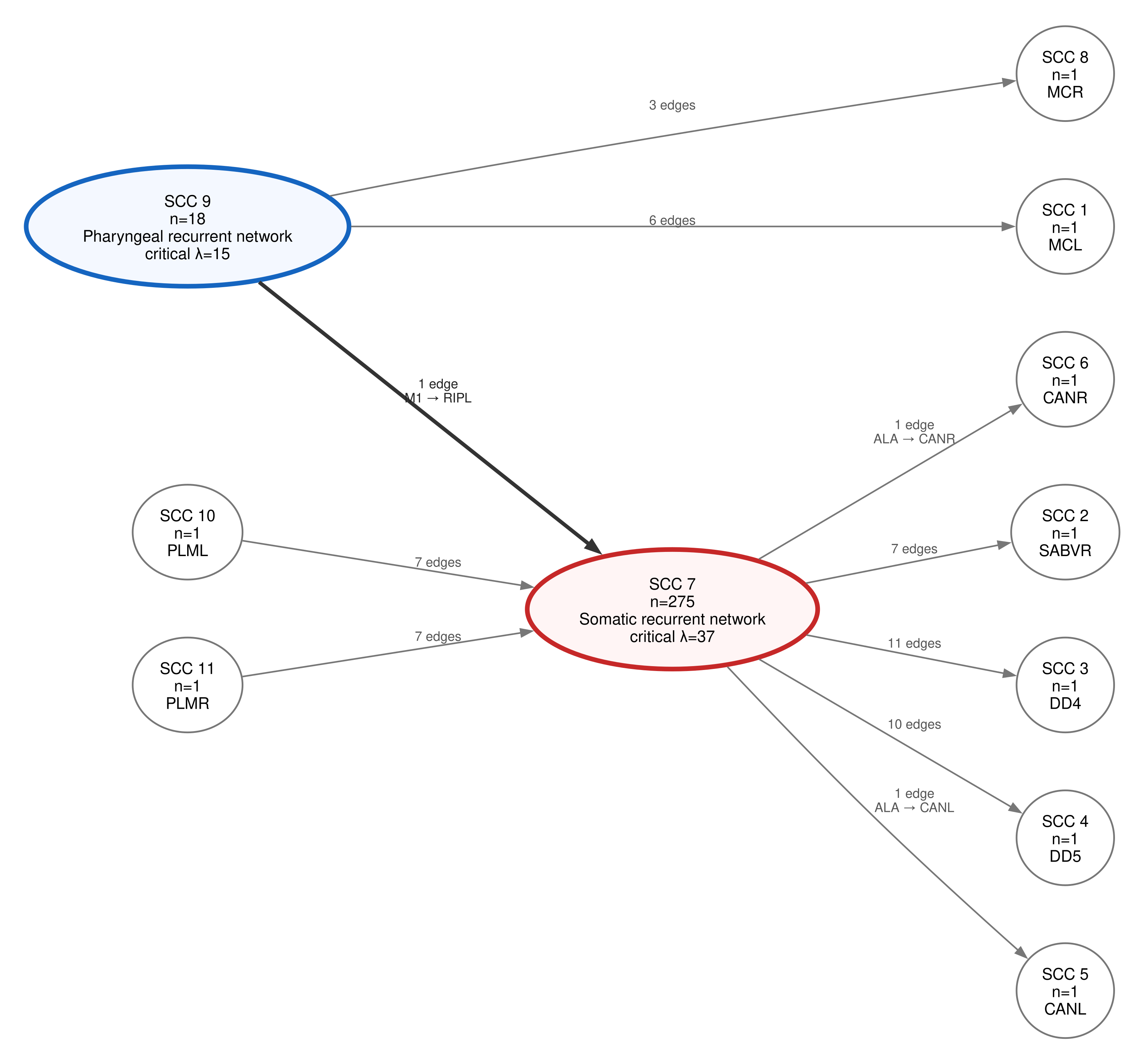}
    \caption[Condensation graph of the chemical-synapse connectome]{Condensation graph of the chemical-synapse connectome. Nodes represent strongly connected components and arrows indicate directed chemical-synaptic connectivity. SCC9 and SCC7 contain the critical classes associated with $\lambda=15$ and $\lambda=37$, respectively. Edge labels indicate the number of neuronal connections between components, with single-edge bridges identified explicitly.}
    \label{fig:synaptic_condensation}
\end{figure}

The 302 neurons separate into 11 SCCs, shown through an extended version of the condensation digraph in Figure~\ref{fig:synaptic_condensation}. The decomposition is highly uneven: SCC7 contains 275 neurons, approximately 91\% of the complete network, SCC9 contains 18 neurons, and the remaining nine SCCs consist of single neurons. This organization closely reproduces the bow-tie architecture previously described for the chemical connectome of \textit{C. elegans}, characterized by a large strongly connected core surrounded by a much smaller directed periphery~\cite{wright2019central}. At the same time, the decomposition recovers the major anatomical division of the nervous system: SCC7 contains the somatic network, whereas SCC9 contains 18 neurons of the pharyngeal nervous system and, together with the corresponding singleton components, accounts for the established separation into 282 somatic and 20 pharyngeal neurons~\cite{cook2019,cook2020pharynx}. The condensation structure therefore captures both the global bow-tie topology of the synaptic connectome and the anatomical separation between its somatic and pharyngeal systems.

SCC9 provides a particularly clear view of the organization of the pharyngeal nervous system. It contains 18 of its 20 neurons and receives no chemical-synaptic connections from any other SCC. Of its three outgoing branches, two terminate in MCL and MCR, the bilateral MC motor neurons, thereby completing the 20-neuron pharyngeal system. MC neurons are cholinergic drivers of rapid pharyngeal pumping~\cite{raizen1995mc} and project extensively to non-neuronal pharyngeal targets, including the marginal cells~\cite{cook2020pharynx}. As these effector connections are not represented in the neuron-to-neuron synaptic matrix, MCL and MCR appear as terminal components in the condensation digraph. The graph therefore captures a recurrent pharyngeal core together with its final neuronal outputs towards the structures responsible for executing feeding.

The third branch is qualitatively different because it connects the pharyngeal recurrent network to the large somatic SCC7. Remarkably, this transition is supported by a single chemical edge, M1$\rightarrow$RIPL. This result agrees with the unusual anatomical position of the RIP neurons at the interface between the pharyngeal and somatic nervous systems. The two systems have historically been described as largely independent and coupled primarily through RIP--pharyngeal electrical connections; the more recent reconstruction of the pharyngeal connectome additionally identified a previously unreported chemical synapse from M1 onto RIP~\cite{cook2020pharynx}. In the chemical matrix analysed here, M1$\rightarrow$RIPL is therefore the unique directed chemical bridge by which the recurrent pharyngeal SCC reaches the recurrent somatic SCC.

SCC7 forms the central recurrent component of the somatic nervous system and contains 275 neurons. It receives three inter-component inputs: the M1$\rightarrow$RIPL bridge from the pharyngeal SCC and independent projections from PLML and PLMR. The latter are the bilateral posterior touch receptor neurons, which detect posterior mechanical stimulation and participate in locomotor responses to body touch~\cite{schafer2015mechanosensory}. Their appearance as two source SCCs is consequently consistent with their peripheral sensory role: neither belongs to the recurrent somatic core, but both project directly into it. Their target sets also overlap strongly, providing bilateral access to a common recurrent network rather than defining two separate downstream systems.

Connectivity then leaves SCC7 through five terminal components. DD4 and DD5 are particularly interpretable in this context. Both belong to the GABAergic D-type motor-neuron class, which innervates body-wall muscles and contributes to dorsoventral inhibition during locomotion~\cite{deng2021inhibition}. Their terminal position again reflects the decision to analyse neuron-to-neuron connectivity: their principal physiological outputs to muscle lie outside the present matrix. SABVR constitutes another singleton downstream branch, while CANL and CANR are each reached by a single connection from ALA. The functions and conventional synaptic roles of the CAN neurons remain substantially less resolved than those of the major sensory and motor classes, so these two sparse branches are retained as anatomical observations without assigning them a stronger functional interpretation. However, their position is particularly notable in light of the gap-junction results presented above. In the gap-junction network, the same neuron pairings form the strongest recurrent structure, with reciprocal ALA--CANL and ALA--CANR couplings of weight 401~\cite{cook2019}. The ALA--CAN motif is therefore prominent in both wired connectivity layers, appearing as exceptionally strong bidirectional electrical coupling in one and directed chemical output in the other. Their repeated association across both connectivity systems highlights a distinctive structural relationship whose physiological role remains unresolved and is worth exploring further.

The SCC decomposition identifies where recurrent paths can exist, but it does not distinguish which recurrent paths dominate once synaptic weights are considered. This is the additional information provided by the max-plus eigenspectrum. 
Two finite eigenvalues emerge from the synaptic network, $\lambda=37$ and $\lambda=15$. They map directly onto the two large SCCs identified above, but their critical classes reveal a much more focused organization. The dominant eigenvalue, $\lambda=37$, belongs to the 275-neuron somatic SCC7, yet its critical class contains only two neurons, RIAR and SMDVL. The complete critical graph consists of the reciprocal pair
\[
\mathrm{RIAR}\xrightarrow{43}\mathrm{SMDVL}
\qquad\text{and}\qquad
\mathrm{SMDVL}\xrightarrow{31}\mathrm{RIAR},
\]
giving the cycle mean
\[
\lambda=\frac{43+31}{2}=37.
\]
Thus, within a recurrent component containing most of the somatic nervous system, the maximum-mean recurrent structure reduces to a two-neuron feedback loop.

The biological identity of this pair gives the spectral result a clear interpretation. RIA interneurons are positioned to combine sensory input with information about head movement, and their axons form reciprocal connections with the SMD head motor neurons~\cite{hendricks2012ria}. Compartmentalized activity in RIA reflects dorsal and ventral head bending and depends on motor input from SMD, while RIA output in turn contributes to steering responses~\cite{hendricks2012ria,liu2018ria}. SMD neurons themselves combine motor and proprioceptive functions involved in coordinating head steering~\cite{yeon2018smd}. The RIAR--SMDVL critical class therefore corresponds to a previously characterized sensorimotor feedback architecture.

The second eigenvalue, $\lambda=15$, is located in the pharyngeal SCC9 and produces a qualitatively different critical structure. Its critical class consists only of M4, through the self-edge
\[
\mathrm{M4}\xrightarrow{15}\mathrm{M4}.
\]
Because this is a cycle of length one, its cycle mean is directly equal to 15. M4 is a cholinergic pharyngeal motor neuron with a central role in feeding: it innervates the posterior isthmus and is required for normal isthmus peristalsis~\cite{trojanowski2014feeding,cook2020pharynx}. The result therefore associates the recurrent spectral mode of the pharyngeal SCC with a neuron already known to occupy a central position in the control of pharyngeal motor behaviour.

These two critical classes provide the anchors for interpreting the corresponding eigendirections. Under the matrix convention used here, a right eigendirection describes how the network is organized in the direction followed by outgoing connectivity from a recurrent spectral class. Its finite coordinates can therefore be interpreted as the finite \emph{downstream domain} associated with that class. Conversely, the left eigendirection describes the complementary \emph{upstream domain}: neurons and components from which the recurrent class can be reached. Right and left eigenvectors thus describe different sides of the same recurrent core.

For $\lambda=37$, the right eigendirection contains 280 finite coordinates. These are precisely the 275 neurons of SCC7 together with all five singleton SCCs reached downstream from it. Starting from the RIAR--SMDVL critical loop, the right eigendirection therefore expands through the entire mutually reachable somatic network and continues into every terminal component accessible from that network.
The left eigendirection of $\lambda=37$ provides the complementary view. It contains 295 finite neurons: the complete 275-neuron somatic SCC, all 18 neurons of the pharyngeal SCC, and the two posterior touch neurons PLML and PLMR. The seven neurons absent from the finite left domain are the terminal components MCL, MCR, SABVR, DD4, DD5, CANL, and CANR, all of which lie downstream of a recurrent component and have no directed route back to the $\lambda=37$ core. For the dominant somatic mode, the right and left eigendirections therefore recover the downstream and upstream organization already suggested by the condensation graph, but now relative specifically to the RIAR--SMDVL spectral core.

The $\lambda=15$ mode exposes the hierarchy between the two recurrent systems even more clearly. Its left eigendirection contains only the 18 neurons of SCC9, explicitly stating that only those neurons have anatomical connections directed to the critical class, M4. In the right eigendirection, the same 18 neurons remain finite and are joined by MCL and MCR, giving a 20-neuron finite domain that corresponds exactly to the pharyngeal nervous system.

The remaining neurons take infinite coordinates of two kinds. PLML and PLMR take the value $\bot$ ($=-\infty$): they are independent source components that cannot be reached from the pharyngeal SCC. The other 280 neurons take the value $\top$ ($=+\infty$); they are exactly the finite right domain of the higher mode $\lambda=37$, that is, the somatic SCC and its five terminal outputs. Through the single bridge M1$\rightarrow$RIPL, the pharyngeal circuit reaches a recurrent component whose own loops have a higher mean weight ($37>15$). Under repeated propagation, the best accumulated weight in the somatic component grows by 37 per step, faster than the 15 per step that defines the pharyngeal mode, so, measured relative to that mode, it increases without bound~\cite{valverde2016}. The $\top$ coordinates therefore mark the downstream region dominated by a stronger recurrent circuit than the one that generated the eigenvector; they encode this hierarchy between recurrent regimes and are not predictions of unbounded neural activity. 

The synaptic connectome can consequently be read at three progressively more specific levels. The SCC decomposition first separates the network into mutually reachable recurrent domains and reveals the directed relations between them. The $\cmaxplus$ eigenvalues then distinguish two weighted recurrent regimes within those domains. Finally, the critical classes reduce each regime to its dominant closed path: the M4 self-loop in the pharyngeal system and the RIAR--SMDVL sensorimotor feedback loop in the somatic system. Right and left eigendirections extend these compact spectral cores towards their downstream and upstream neuronal domains, respectively. Together, these representations transform a dense 302-neuron chemical connectome into an interpretable hierarchy of recurrent cores, directed inter-system transitions, sensory inputs, and terminal outputs while preserving the identity of the individual neurons supporting each structure.

\subsubsection{Monoaminergic connectome}
\label{sec:maxplus_monoamine}

The monoaminergic layer displays a highly asymmetric recurrent organization, markedly different from both wired connectivity layers examined above. Whereas the chemical-synapse connectome is dominated by a large recurrent somatic core and the gap-junction network places almost the entire nervous system within a single SCC, the monoaminergic condensation graph contains 286 SCCs, almost all of them singletons. The only sizeable recurrent component is SCC268, which contains 17 neurons. This compact core receives connections from only two upstream SCCs but projects to 267 downstream SCCs, producing a strongly divergent architecture in which a small recurrent population broadcasts monoaminergic influence across most of the nervous system. The remaining 16 singleton SCCs (the AFD, ASI, AWA, AWB and PHC pairs, PQR, PDA, VC1--3 and VC6) have no monoaminergic connection to or from the rest of the network in this representation.

The biological composition of SCC268 closely reflects the known organization of monoaminergic signalling in the hermaphrodite nervous system. Sixteen of its seventeen neurons correspond to the principal monoamine-releasing classes: the serotonergic ADF, NSM, and HSN pairs; the dopaminergic CEP and PDE neurons; the tyraminergic RIM pair; and the octopaminergic RIC pair~\cite{bentley2016,wang2024}. The remaining neuron, I5, has been associated with less conventional and comparatively uncertain monoaminergic, particularly serotonergic, signalling in the updated neurotransmitter atlas~\cite{wang2024}. SCC268 therefore concentrates almost the entire canonical monoaminergic broadcasting system within a single recurrent component.

The two remaining dopaminergic neurons, ADEL and ADER, occupy a distinctive upstream position. They form the only two singleton SCCs feeding into SCC268 and, together with the 17-neuron recurrent component, constitute the complete finite domain of the left eigendirection. The left eigenvector consequently contains only 19 finite coordinates. In the opposite direction, the right eigendirection contains 284 finite neurons: the 17 neurons of SCC268 together with the 267 downstream singleton SCCs reached from it. 

\begin{figure}[h]
    \centering
    \includegraphics[width=\textwidth]{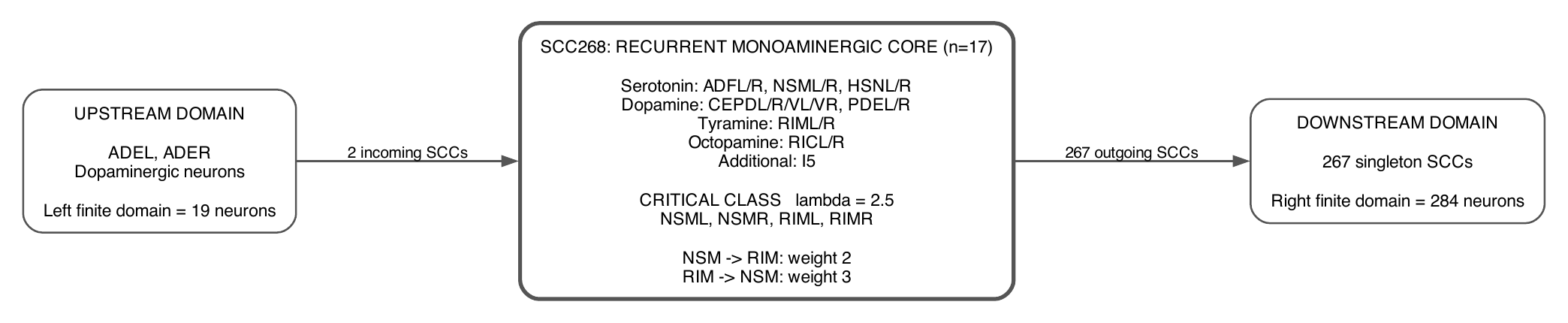}
    \caption[Structural and spectral organization of the monoaminergic network]{Structural and spectral organization of the monoaminergic network. ADEL and ADER form the restricted upstream domain feeding the 17-neuron recurrent monoaminergic SCC, which reaches 267 downstream singleton SCCs.}
    \label{fig:monoamine_architecture}
\end{figure}

Figure~\ref{fig:monoamine_architecture} summarizes this organization and places the spectral structure within the biological composition of the network.

This strong asymmetry provides a spectral representation of the broadcasting character of monoaminergic signalling. A very restricted upstream domain converges on the recurrent aminergic population, whereas paths leaving this core extend across most of the neuronal network. This organization closely parallels the topology described by Bentley et al., who characterized the monoaminergic connectome as a highly disassortative, star-like network in which a small set of interconnected monoamine-releasing neurons acts as hubs projecting towards a much larger population of receptor-expressing targets~\cite{bentley2016}. 

The monoaminergic layer contains a single finite eigenvalue, $\lambda=2.5$. Its critical class contains only NSML, NSMR, RIML, and RIMR. In this matrix, a directed edge represents potential monoaminergic signalling from a neuron releasing a given monoamine to a target expressing compatible receptors, and its weight counts the supported monoamine--receptor relations between them. Every NSM neuron participates in critical connections with both RIM neurons: the NSM-to-RIM edges have weight 2, whereas the RIM-to-NSM edges have weight 3. The critical graph therefore contains four equivalent length-two cycles of the form
\[
\mathrm{NSM}\xrightarrow{2}\mathrm{RIM}
\xrightarrow{3}\mathrm{NSM},
\]
each with cycle mean
\[
\frac{2+3}{2}=2.5.
\]
Thus, $\lambda=2.5$ expresses the maximum mean number of supported monoamine--receptor relations along a recurrent cycle in this network.

The identity of this critical class is biologically meaningful. NSM neurons are serotonergic neurosecretory neurons with a central role in the regulation of feeding, whereas RIM neurons constitute the principal tyraminergic neuronal pair. Experimental work has previously placed NSM and RIM within a central monoaminergic circuit regulating pharyngeal pumping and behavioural state. In the ``flip-flop'' circuit described by Li et al., serotonergic signalling from NSM promotes the feeding state, while tyraminergic and octopaminergic signalling involving RIM and RIC contributes to the opposing behavioural state; NSM and RIM activity also responds in opposite directions to attractive and aversive sensory cues~\cite{li2012flipflop}. The max-plus analysis therefore isolates, from the broader recurrent monoaminergic SCC, an NSM--RIM interaction that closely corresponds to an experimentally characterized regulatory circuit.

Overall, the monoaminergic analysis provides a particularly direct correspondence between max-plus spectral organization and the established architecture of extrasynaptic monoaminergic neuromodulation. A compact recurrent SCC gathers almost the complete set of monoamine-releasing neurons, the left and right eigendirections recover the strong asymmetry between a restricted upstream domain and widespread downstream broadcasting, and the critical graph further reduces this recurrent population to the experimentally supported NSM--RIM regulatory interaction. The three levels of analysis therefore distinguish the global broadcasting architecture, its recurrent monoaminergic core, and the specific weighted recurrence that dominates within that core.

\subsubsection{Neuropeptidergic connectome}
\label{sec:maxplus_neuropeptide}

\begin{figure}[ht]
    \centering
    \includegraphics[width=0.94\textwidth]{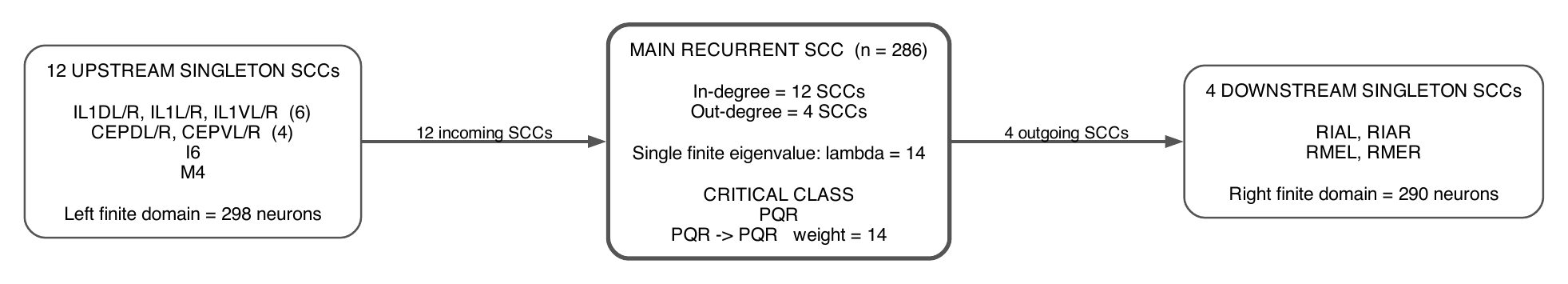}
    \caption[Structural and spectral organization of the short-range neuropeptidergic network]{Structural and spectral organization of the short-range neuropeptidergic network. Twelve singleton SCCs feed into the main 286-neuron recurrent component, while four singleton SCCs are reached downstream from it.}
    \label{fig:neuropeptide_architecture}
\end{figure}

The neuropeptidergic layer recovers a broadly recurrent organization, in marked contrast with the highly fragmented monoaminergic network. Its 302 neurons are partitioned into only 17 strongly connected components, with a single dominant SCC containing 286 neurons. The remaining 16 neurons form singleton SCCs arranged around this recurrent domain: 12 lie upstream of it and four lie downstream. Figure~\ref{fig:neuropeptide_architecture} summarizes this organization together with the corresponding $\cmaxplus$ critical structure.

The external SCCs show a notable organization by neuronal class. The upstream domain contains the complete six-neuron IL1 class, all four CEP neurons, and the individual neurons I6 and M4. Conversely, the downstream domain consists of the bilateral RIA and RME pairs. This class-wise separation is consistent with the cell-type-specific combinations of neuropeptides and receptors from which the network is constructed~\cite{ripollsanchez2023}. 

Several aspects of this peripheral organization are also biologically coherent. IL1 neurons constitute a peptidergically active sensory--motor class, while M4 combines its established pharyngeal motor function with neuroendocrine peptide release. The four CEP neurons likewise remain together as a complete neuronal class on the upstream side. On the downstream side, the paired organization of RIA and RME again follows neuronal identity. Importantly, RIA itself is known to release neuropeptides, so its terminal position here does not imply absence of peptidergic output; it indicates that those outputs do not generate a directed path back into the main recurrent SCC within the particular short-range ligand--receptor network analysed. The separation of these 16 neurons therefore appears structured by molecular connectivity.

Despite the size of the recurrent component, its dominant weighted recurrence is maximally localized. The neuropeptidergic layer contains a single finite eigenvalue, $\lambda=14$, whose critical class consists only of PQR. The corresponding critical cycle is the self-loop
\[
\mathrm{PQR}\xrightarrow{14}\mathrm{PQR}.
\]
In this network, an edge represents a potential interaction from a neuron releasing a neuropeptide to a neuron expressing a compatible receptor under the short-range connectivity model, and its weight counts the supported neuropeptide--receptor relations. The PQR self-loop therefore represents 14 distinct supported ligand--receptor routes for potential autocrine signalling.

This result has direct support in the original neuropeptidergic connectome analysis. Autocrine signalling is widespread in the predicted network, and PQR was identified among the sensory neurons with particularly high diversity of neuropeptidergic self-interactions~\cite{ripollsanchez2023}. PQR belongs to the oxygen-sensing circuit together with AQR and URX and exhibits tonic activity related to environmental oxygen levels. The emergence of its self-loop as the maximum-mean cycle therefore isolates an already recognized form of unusually rich autocrine signalling from within an otherwise extremely dense recurrent network.

The eigendirections show how the PQR critical structure is embedded within the broader neuropeptidergic network. The left eigendirection contains 298 finite coordinates, including the 286 neurons of the main recurrent SCC together with the 12 upstream singleton components, all of them capable of signalling towards PQR; whereas the right eigendirection contains 290 finite coordinates, comprising the same recurrent SCC and the four downstream singletons, all of them capable of receiving signalling from PQR. The large overlap between both eigendirections reflects the extensive mutual reachability of the neuropeptidergic core, while their small difference is confined to the directional periphery.

Overall, the neuropeptidergic network is characterized by widespread recurrent connectivity, with 286 of the 302 neurons belonging to a single strongly connected component. The few neurons outside this component appear as coherent neuronal classes occupying either upstream or downstream positions. Within this broadly recurrent architecture, the max-plus analysis identifies a much more specific organizing structure: the autocrine PQR self-loop, supported by 14 neuropeptide--receptor relations, defines the maximum-mean recurrence of the network. This separates the extensive potential for reciprocal neuropeptidergic communication across the nervous system from the particular interaction that dominates its weighted recurrent organization.

\subsubsection{Functional signal-propagation network}
\label{sec:maxplus_functional}

The final layer moves from anatomically reconstructed connectivity to experimentally measured signal propagation. Randi et al. combined single-neuron two-photon optogenetic stimulation with whole-brain calcium imaging to quantify how activation of an upstream neuron affects activity in downstream neurons~\cite{randi2023}. The resulting functional network therefore has a different interpretation from the preceding layers: a directed edge represents a statistically supported downstream calcium response following stimulation of another neuron, and its weight is the mean post-stimulus response amplitude. The present matrix contains 188 neurons represented in the functional atlas.

The network contains 42 strongly connected components, but only one contains more than a single neuron. This main recurrent component comprises 147 neurons, while the remaining 41 SCCs are singletons. Eleven singleton SCCs lie upstream of the recurrent component and 19 lie downstream, whereas the remaining 11 fall outside both corresponding eigendomains. The functional network therefore combines a large recurrent signal-propagation domain with a comparatively sparse peripheral organization.

\begin{figure}[ht]
    \centering
    \includegraphics[width=0.96\textwidth]{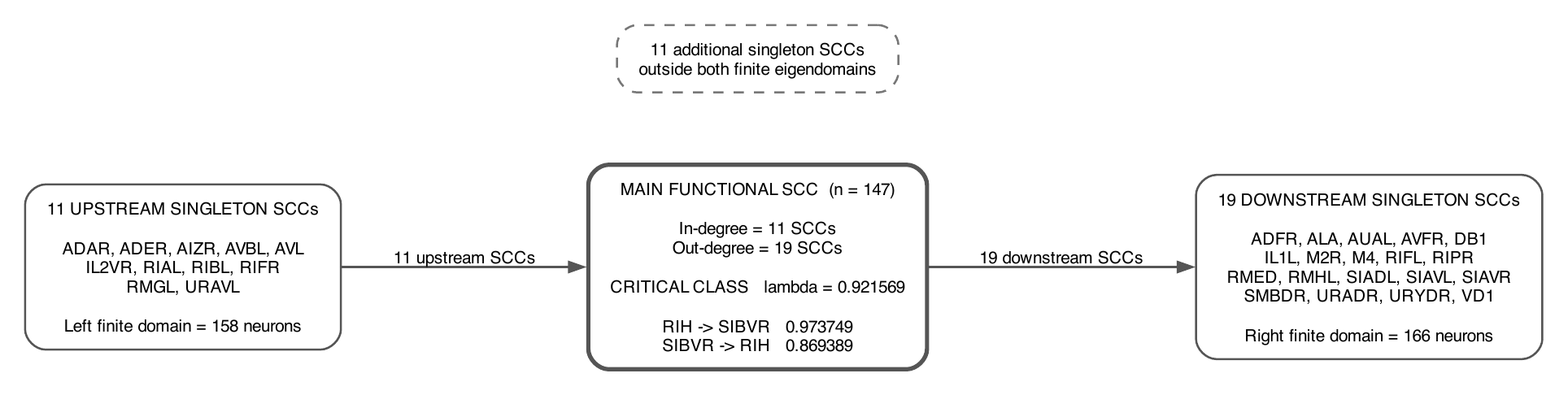}
    \caption[Structural and spectral organization of the functional signal-propagation network]{Structural and spectral organization of the functional signal-propagation network. Eleven singleton SCCs can reach the main 147-neuron recurrent component, which in turn reaches 19 downstream singleton SCCs. Eleven additional singleton SCCs lie outside both finite eigendomains.}
    \label{fig:functional_architecture}
\end{figure}

Figure~\ref{fig:functional_architecture} summarizes this structure together with the dominant $\cmaxplus$ recurrence.

The unique finite eigenvalue is $\lambda=0.921569$, and its critical class contains only RIH and SIBVR. Optogenetic activation of RIH produces a downstream response of $0.973749$ in SIBVR, while stimulation of SIBVR produces a response of $0.869389$ in RIH. These two edges form the critical cycle
\[
\mathrm{RIH}\xrightarrow{0.973749}\mathrm{SIBVR}
\xrightarrow{0.869389}\mathrm{RIH},
\]
with mean
\[
\lambda =
\frac{0.973749+0.869389}{2}
=0.921569.
\]
In this layer, the eigenvalue therefore has a particularly direct interpretation: it is the maximum mean post-stimulus calcium-response amplitude sustained around this recurrent functional cycle. It measures experimentally observed signal propagation.

The identity of the critical pair places the dominant functional recurrence at an interesting interface between neuromodulation and motor circuitry. RIH does not synthesize serotonin, but takes up extracellular serotonin through MOD-5/SERT and contributes to regulating its spatial and temporal availability, with consequences for locomotor adaptation to food availability~\cite{jafari2011serotonin}. SIBVR belongs to the cholinergic SIB sublateral motor-neuron class, whose longitudinal processes and neuromuscular outputs place these neurons within the circuitry coordinating head and body movements~\cite{cook2019}. The reciprocal RIH--SIBVR responses therefore connect a neuron involved in state-dependent serotonergic regulation with a component of the somatic motor system. Although the functional atlas does not establish the pathway responsible for this interaction, its emergence as the maximum-mean cycle suggests a particularly strong functional coupling between modulatory and motor domains.

The eigendirections place the RIH--SIBVR critical cycle within the wider functional propagation network. The left eigendirection contains 158 finite neurons, comprising the 147-neuron recurrent SCC together with the 11 upstream singleton components, whereas the right eigendirection contains 166 finite neurons, including the same recurrent SCC and the 19 downstream singletons reached from it. This organization indicates that the dominant functional recurrence is embedded within a broadly interconnected propagation core, with a modest but clearly asymmetric shell of neurons positioned before and after that recurrent domain.

Overall, the functional network reveals extensive recurrent propagation across the measured head nervous system, with 147 of the 188 neurons participating in a common strongly connected domain. Within this broad functional recurrence, the $\cmaxplus$ spectrum identifies a highly localized dominant cycle between RIH and SIBVR. The result complements the anatomical and neuromodulatory layers by locating the maximum-mean recurrence directly in experimentally observed neural responses.

Across the five networks, recurrent domains range from the compact monoaminergic core to the almost system-wide electrical and neuropeptidergic components. Within these domains, the maximum-mean cycles consistently localize weighted recurrence to small neuronal structures. The chemical and monoaminergic results recover interactions associated with established sensorimotor and regulatory circuits, while the electrical and functional results identify specific associations for further physiological investigation. These comparisons concern the organization and neuronal identity of the recovered structures, with spectral magnitudes interpreted in the measurement units of each dataset.

\subsection{Comparison with conventional connectivity analyses}\label{sec:comparison}
 
The \emph{C.~elegans} connectome has been characterized with a broad repertoire of network-analysis methods, including graph-theoretic measures such as degree, path length, efficiency, and reciprocity~\cite{rubinov2010,varshney2011}, community detection~\cite{newman2006,blondel2008,girvan2002,onuchin2023,emmons2024}, hub and rich-club analyses~\cite{colizza2006,towlson2013,uzel2022,lin2024}, and spectral and low-dimensional representations~\cite{varshney2011,petrovic2019,jolliffe2016}. Each answers a different question about network organization. Here we ask what the added value of the max-plus framework is with respect to all those consolidated techniques. We compare it first with its closest conventional counterpart, a linear spectral analysis of the same five matrices, and then with community detection and rich-club analysis. For the latter we rely on the published literature and, as a basic baseline applied to the same four anatomical and molecular layers, on a previous bachelor's thesis from our group~\cite{casado2023}. We cite it explicitly as a basic baseline: it applies standard, off-the-shelf methods in the four connectomes. %and is not intended as a state-of-the-art comparison.
 
\subsubsection{Linear spectral analysis}
 
Linear and max-plus spectral analyses sit at opposite ends of the spectrum of network communication models introduced above~\cite{seguin2023}. The linear eigenproblem $Av=\lambda v$ adds the contributions of all incoming connections, so that $(A^{k})_{ij}$ sums the weights of \emph{all} walks of length $k$ from $j$ to $i$, as in diffusion. The max-plus eigenproblem keeps only the strongest one, $(A\mpow{k})_{ij}$, as in routing. We applied the standard linear toolkit to each matrix, with the same orientation as in the max-plus analysis: eigendecomposition, eigenvector centrality (leading right and left eigenvectors, i.e.\ in- and out-centrality)~\cite{bonacich1987}, Katz centrality~\cite{katz1953}, and spectral clustering of the symmetrized network~\cite{fiedler1973,vonluxburg2007}.
 
The two analyses agree on the coarse organization. In the gap-junction network, ALA, CANR, and CANL are the three most central neurons, matching the ALA--CANL/R critical class (Section~\ref{sec:maxplus_gapjunction}). In the chemical-synapse network, RIAR is among the six most central neurons in both directions, and spectral clustering separates the 20 pharyngeal neurons from the somatic network, as the SCC decomposition does. The leading linear eigenvectors also extend over the same neurons as the max-plus finite domains (for example, 280 and 295 neurons for the somatic synaptic mode), because both depend on which neurons can reach which.
 
Beyond this, the linear description loses three kinds of information that max-plus retains. First, it summarizes each layer through one dominant mode, so secondary recurrent circuits disappear. The pharyngeal M4 self-loop ($\lambda=15$) has no counterpart among the leading linear modes: M4 ranks 285th and 292nd of 302 neurons in in- and out-centrality, and the linear mode that weights it most is only the 18th. The independent IL2DL--URADL electrical pair ranks 297th--298th. Second, linear centrality largely follows the number and strength of connections: in the neuropeptidergic network its correlation with node strength is 0.98--0.99, and PQR ranks only 24th in in-centrality, whereas max-plus singles out its autocrine loop as the dominant recurrence. In the functional network, RIH and SIBVR rank between 61st and 119th, and their interaction appears only in the fourth linear mode. Third, beyond the leading mode, the linear spectrum is hard to read: in the chemical-synapse and functional networks more than a hundred eigenvalues exceed 10\% of the spectral radius, most are complex, their eigenvectors mix many neurons with positive and negative coordinates, and the eigenvector bases are so ill-conditioned (condition numbers above $10^{18}$ in all four directed layers) that small changes in the data can alter them substantially~\cite{trefethen2005}. The max-plus structures, by contrast, are exact objects read directly from the critical graph and the condensation graph.
 
\subsubsection{Biological meaning of the spectral quantities}
 
The main difference between the two descriptions is, thus, how directly their outputs can be read in biological terms. Because max-plus adds weights along a route and selects between routes, every quantity it returns is expressed in the units of the original data and is attained by identifiable connections. The eigenvalue is the mean weight per connection of a specific closed route: 37 is the average of the 43 and 31 reconstructed contacts of the RIAR--SMDVL loop, 14 is the number of supported ligand--receptor relations of the PQR self-loop, and 0.92 is the mean calcium response propagated around the RIH--SIBVR cycle. The critical graph names the neurons and connections responsible for that value, and the eigenvector coordinates give the best accumulated weight linking each neuron to the critical circuit.
 
The linear quantities have no such reading. The spectral radius is the growth rate of sums of products of weights over all walks; it is not attained by any circuit, and a product of electron-microscopy section counts, or a sum of thousands of such products, is not a measurable quantity. Accordingly, the linear spectral radii (106.1, 716.5, 16.3, 358.2, and 1.41 for chemical synapses, gap junctions, monoamines, neuropeptides, and functional propagation) exceed the max-plus eigenvalues by factors of about 1.5 to 26, which reflect the density of parallel routes rather than the strength of any interaction. Linear eigenvector coordinates are relative amplitudes of a collective mode and, beyond the leading one, can be negative or complex, which no anatomical contact count, receptor count, or neuronal state can be.
 
These differences follow from the space in which each method represents the network. A linear analysis places neurons in a Euclidean vector space, where patterns can only be combined by adding and rescaling them. To express that one part of the network is dominated by another, or that a neuron cannot be reached from a circuit, it must use cancellations between positive and negative or complex components. The chemical-synapse network shows this directly. The pharyngeal circuit reaches the somatic core, whose recurrence is stronger; the linear mode closest to the pharyngeal circuit encodes this with a change of sign across the somatic network, whereas the max-plus eigenvector encodes it as an order relation, assigning $\top$ to the dominating region, finite values to the pharyngeal domain, and $\bot$ to the unreachable neurons (Section~\ref{sec:maxplus_synaptic}). More generally, the structural results of the max-plus analysis are statements of order or inclusion: which neurons belong to a recurrent domain, which domains lie upstream or downstream of others, and which circuit dominates each region. The fact that the connectivity data are described exactly and economically in these terms, and only through cancellations in Euclidean terms, supports the proposal that the states supported by these networks are better represented in an ordered, lattice-structured space than in a Euclidean one.
 
\subsubsection{Community detection and rich clubs}
 
Community detection and rich-club analysis address the segregation and integration of the network rather than its recurrent organization, and have recovered meaningful structure in \emph{C.~elegans}. The most comprehensive recent analysis of the wired connectome partitioned it into ten communities that largely correspond to sensory-to-motor pathways and identified integrative interneuron hubs~\cite{emmons2024}; non-backtracking spectral methods recover a similar partition~\cite{onuchin2023}. Rich-club analysis identifies a densely interconnected core of command interneurons that control locomotion~\cite{towlson2013,uzel2022}, and the monoaminergic network shows a star-like organization around interconnected broadcasting hubs~\cite{bentley2016}.
 
Our results agree with these descriptions and add to them. RIA, one of the most convergent premotor interneurons in~\cite{emmons2024}, is part of the RIAR--SMDVL critical cycle. The monoaminergic broadcasting hubs correspond to the 17-neuron recurrent core identified here; in the basic baseline, 14 of the 16 neurons of the monoaminergic out-degree rich club belong to this core~\cite{casado2023}. Max-plus adds three things to such hub sets: it places ADEL and ADER upstream of the core rather than within it, it adds RIC and I5 to the core, and it identifies NSM--RIM, a characterized circuit regulating feeding and behavioral state~\cite{li2012flipflop}, as the dominant recurrence inside it. In the gap-junction and neuropeptidergic layers, the baseline found either no distinct rich club or very large ones of 30 to 50 neurons, whereas max-plus isolates a single critical circuit in each. These comparisons with the baseline are qualitative, because it used earlier versions of the synaptic and neuropeptidergic data and a 300-neuron connectome that excluded CANL and CANR.
 
These differences again come from what each method can represent. Community partitions are usually computed on undirected, and sometimes unweighted, graphs~\cite{onuchin2023,emmons2024}, and the most complete analysis merged chemical and electrical contacts into a single graph~\cite{emmons2024}. In the basic baseline, this grouped pharyngeal and anterior head neurons into a single ``brain'' community~\cite{casado2023}, whereas the directed SCC decomposition separates the pharyngeal and somatic recurrent systems and shows that they are joined by a single chemical edge, M1$\rightarrow$RIPL. Even when direction and weight are retained~\cite{sohn2011}, modularity does not say which modules are recurrent, how they are ordered along the direction of propagation, or which connections dominate within them. These three aspects are given by the SCC decomposition, the condensation graph, and the critical graph, respectively.
 
Overall, conventional analyses and max-plus analysis agree on the coarse organization of the \emph{C.~elegans} connectome: the separation between pharyngeal and somatic systems, the broadcasting character of the monoaminergic layer, the prominence of ALA--CAN electrical coupling, and the hub role of monoamine-releasing neurons. Conventional methods remain the appropriate tools for describing modules, hubs, and global spectral properties. For identifying the recurrent circuits that organize each communication layer and how they are embedded in the network, max-plus analysis offers four advantages. First, and most importantly, its outputs keep a direct biological meaning: eigenvalues, critical cycles, and eigenvector coordinates are expressed in the units of the data and are attained by named neurons and connections, whereas the linear spectral radius and eigenvectors, centrality scores and communities, are aggregates of many routes or neurons whose biological meaning is lost. Second, it resolves coexisting recurrent circuits, such as the pharyngeal and somatic loops, instead of a single dominant mode. Third, it preserves the direction of propagation, placing each circuit between its upstream and downstream neuronal domains. Fourth, it describes all of these structures through relations of order and inclusion rather than through cancellations between signed components, consistent with an ordered rather than Euclidean organization of the states these networks support.

\section{Discussion}\label{sec12}

A central challenge in connectomics is understanding how different mechanisms of neuronal communication shape the recurrent organization through which neural activity can persist and evolve. This study advances that goal through three connected contributions: a time-dependent model of the functional connectome, a generalized max-plus spectral method for characterizing recurrent organization, and an empirical analysis that links the resulting structures to established neurobiology while identifying new targets for investigation. Together, these contributions move from the description of individual connections towards an interpretable account of the circuits and neuronal domains that organize network-wide interactions.

The first contribution is a model in which the functional connectome emerges from the changing contributions of superposed communication layers. Electrical coupling, chemical synapses, monoaminergic and neuropeptidergic signaling, and potentially other mechanisms operate over different timescales and can therefore reshape the effective network as their contributions develop or decay. Representing their composition through a time-dependent max-plus operator connects this biological heterogeneity to a concrete account of network reorganization: the engagement of a signaling layer may create new routes between neuronal groups, change which feedback cycles dominate within an existing recurrent domain, or modify the pathways linking those cycles to the rest of the system. Importantly, critical circuits may change even when the overall pattern of structural connectivity remains unchanged, and they depend only on the relative engagement of the layers. The model thus provides a way to investigate how shifts in the relative contribution of neuronal communication mechanisms could support different configurations of internal brain operation. We have presented it here in the simplified form needed to interpret the empirical analyses; its full mathematical development, including the formal treatment of layer composition and the organization of the resulting state space as a hierarchy of lattices, is given in a companion paper~\cite{companion}.

The second contribution is methodological. By applying generalized $\cmaxplus$ spectral analysis to directed, weighted, and potentially reducible connectomes, we obtain a coherent description of recurrent organization at complementary scales. Strongly connected components establish the domains of possible recurrence; the critical graph identifies the cycles that dominate according to their mean edge weight; and fundamental left and right eigendirections place these recurrent structures within their upstream and downstream neuronal domains. The all-eigenvectors construction is particularly important for neuronal networks, where distinct recurrent components may coexist and be connected asymmetrically. Retaining their full spectral organization yields a representation that preserves both the specific neurons participating in critical interactions and the larger architecture within which those interactions are embedded. This combination provides a biologically interpretable bridge between local feedback circuits and system-level connectivity.

The third contribution is empirical, and it provides strong support for the usefulness of this representation. Across connectivity layers with markedly different architectures, max-plus analysis recovers structures that correspond to well-established features of the \emph{C.~elegans} nervous system. The synaptic decomposition reflects the separation between the somatic and pharyngeal systems, while the RIAR--SMDVL critical cycle is consistent with the known sensorimotor feedback between RIA interneurons and SMD motor neurons~\cite{hendricks2012ria,liu2018ria}. In the monoaminergic layer, the restricted recurrent core and widespread downstream organization recover the characteristic broadcasting architecture of extrasynaptic signaling~\cite{bentley2016}; within that core, the NSM--RIM critical interactions correspond to a previously characterized circuit involved in feeding and behavioral-state regulation~\cite{li2012flipflop}. Likewise, the PQR critical self-loop in the neuropeptidergic network highlights a neuron already recognized for its unusually rich predicted autocrine signaling~\cite{ripollsanchez2023}. Particularly significant is that these correspondences emerge from the same spectral formulation despite the different biological mechanisms and network structures involved. Their convergence with established findings demonstrates the capacity of the method to extract meaningful neuronal organization and provides a compelling basis for interpreting its less-characterized results.

Indeed, the framework also brings attention to recurrent relationships whose roles are not yet comparably understood. The exceptionally prominent ALA--CANL/R electrical interactions, together with the corresponding ALA--CAN connections in the synaptic network, reveal a distinctive association across two communication layers. In the experimentally measured functional network, the RIH--SIBVR critical cycle identifies a strong reciprocal propagation relationship linking a neuron involved in serotonergic regulation with a component of motor circuitry. These results are valuable precisely because the same analysis that recovers recognized anatomical and functional organization also singles out these less-explored associations. They offer focused candidates for physiological investigation, including targeted perturbations and brain-wide recordings to determine how their activity contributes to recurrent propagation and behavioral-state organization.

Several limitations also point to the next steps. The anatomical and molecular layers encode potential interactions whose weights are proxies for physiological strength and do not distinguish excitation from inhibition, which could be represented explicitly through the dual min-plus operation. The neuropeptidergic analysis uses the conservative short-range model, and no neuron-to-neuron hormonal map is yet available. Above all, reconstructing the composite operator of Eq.~\eqref{eq:composition} will require bringing the layers onto a common scale, so that the predicted reorganization of critical circuits can be tested against whole-brain recordings across behavioral states.

Taken together, this work proposes a different way of reading a connectome: not as a collection of nodes to rank or modules to partition, but as a set of dominant feedback circuits, each embedded between the neurons that can drive it and the neurons it can reach. Max-plus analysis makes this reading exact and keeps it biological, because every structure it returns is expressed in the units of the original measurements and attained by identified neurons and connections. Combined with a model in which communication mechanisms reshape these circuits over time, it turns the static wiring diagram into a hypothesis about how internal states are sustained and switched, and it suggests that those states are organized by order rather than by Euclidean geometry. As connectomes and brain-wide recordings become available for more species and conditions, a framework that preserves direction, mechanism, and biological meaning will be essential to move from mapping nervous systems to understanding how they operate.

\section{Methods}\label{sec11}

\subsection{Connectivity datasets}
\label{sec:connectivity_datasets}

The connectivity analysis is based on five complementary network descriptions of the adult hermaphrodite \emph{C.~elegans} nervous system: electrical gap junctions, chemical synaptic connectivity, monoaminergic signaling, neuropeptidergic signaling, and experimentally measured functional signal propagation. All five datasets are expressed over identified neurons, which makes it possible to relate network-level structures back to named biological components. They represent different aspects of neuronal interaction and therefore differ in directionality, density, spatial range, experimental origin, and interpretation of their edge weights~\cite{dvali2025,venkatesh2025}. For this reason, they are retained as separate networks throughout the analysis.

\subsubsection{Electrical gap-junction connectome}
\label{sec:data_gap}

Gap junctions are channels between the cytoplasms of coupled cells that allow ionic currents to pass directly between them, supporting rapid electrical communication. Electrical connectivity was extracted from the whole-animal adult-hermaphrodite connectome reconstructed by Cook et al.~\cite{cook2019}. The corrected July 2020 adjacency matrices distributed through WormWiring were used~\cite{wormwiring2020}. These matrices incorporate corrections introduced after publication and provide a standardized representation of the adult hermaphrodite wiring diagram.

Only neuronal nodes were retained, yielding the complete set of 302 neurons. Non-neuronal cells and neuromuscular contacts were excluded so that the resulting matrix represents neuron-to-neuron connectivity exclusively. Gap junctions describe direct electrical coupling between cells and are represented in the dataset as symmetric connections, so that the resulting adjacency matrix is undirected at the level of graph topology.

The weights correspond to the values reported in the corrected gap-junction adjacency matrix. They represent the number of serial electron-microscopy sections over which connectivity is observed, incorporating the number and anatomical extent of the reconstructed contacts~\cite{cook2019}. Both rows and columns index the same neuronal set, and the symmetry of the matrix implies that
\begin{equation}
A^{(G)}_{ij}=A^{(G)}_{ji}.
\label{eq:methods_gap_symmetry}
\end{equation}
Consequently, exchanging rows and columns preserves the electrical connectivity representation.

\subsubsection{Chemical synaptic connectome}
\label{sec:data_synaptic}

The chemical-synapse network was obtained from the same corrected Cook et al.\ reconstruction and WormWiring release~\cite{cook2019,wormwiring2020}.

Chemical synapses define a directed weighted graph: an edge from neuron $i$ to neuron $j$ represents a reconstructed chemical synaptic relation from the presynaptic to the postsynaptic neuron, and its weight is given by the corresponding value in the corrected adjacency matrix. As for the gap-junction network, only neuronal nodes were retained, yielding a matrix over the 302 neurons of the adult hermaphrodite. The weights describe the reconstructed anatomical contacts.

In the original Cook et al.\ adjacency matrices, rows correspond to presynaptic neurons and columns to postsynaptic neurons. Thus, an entry in row $i$ and column $j$ represents a weighted directed connection $i\rightarrow j$. This orientation is explicitly taken into account during preprocessing before the max-plus analysis.

\subsubsection{Monoaminergic connectome}
\label{sec:data_monoamine}

The monoaminergic network captures potential extrasynaptic interactions mediated by the classical monoamines dopamine, serotonin, tyramine, and octopamine. The layer used here was obtained through the hermaphrodite implementation provided by the \emph{C.~elegans} Connectome Toolbox~\cite{connectometoolbox}.

This implementation integrates several complementary sources. Neuronal transmitter identities are informed by the updated neurotransmitter atlas of Wang et al.~\cite{wang2024}; neuronal identities and the adult-hermaphrodite reference system are aligned with the Cook et al.\ connectome; and monoaminergic receptor-expression relationships derive from the multilayer signaling framework originally assembled by Bentley et al.~\cite{bentley2016}. A directed interaction therefore links a monoamine-producing neuron with a neuron expressing a compatible receptor, independently of whether the two cells are connected by a conventional synapse.

The corresponding connectivity matrices follow a source--target convention in which rows represent the monoamine-producing, or presynaptic-equivalent, neuron and columns represent the receptor-expressing target neuron. An entry in row $i$ and column $j$ therefore represents a weighted directed connection $i\rightarrow j$. Its weight counts the supported monoamine--receptor relationships between the two neurons. The resulting matrix is interpreted as an inferred transmitter--receptor signaling network. Its edges describe potential routes of monoaminergic influence inferred from transmitter and receptor identity, providing a whole-system representation of wireless neuromodulatory communication that can be compared with the wired connectome.

\subsubsection{Neuropeptidergic connectome}
\label{sec:data_neuropeptide}

The neuropeptidergic network was obtained from the system-wide signaling connectome reconstructed by Ripoll-S\'anchez et al.~\cite{ripollsanchez2023}. This resource combines single-cell neuropeptide and receptor expression with experimentally characterized peptide--receptor relationships to infer potential signaling interactions between identified neurons. As in the monoaminergic network, an edge represents a ligand--receptor-compatible communication route.

The updated matrices released by the authors in February 2024 were used. These incorporate revised single-cell expression information for \emph{dmsr-5} and \emph{npr-34}. The resource provides several spatial models that differ in the distance over which peptide signaling is allowed to act. In this study, the short-range model is used as the neuropeptidergic dataset because it constitutes the most spatially conservative reconstruction. The corresponding mid- and long-range versions provide alternative representations in which progressively more distant ligand--receptor interactions are permitted. This distinction reflects the broader spatial and temporal reach of neuropeptide signaling~\cite{watteyne2024}.

The released matrices follow the same source--target orientation used for the monoaminergic representation. Rows correspond to peptide-producing source neurons and columns to neurons expressing compatible receptors. An entry in row $i$ and column $j$ therefore represents a weighted directed connection $i\rightarrow j$, with its weight counting the supported neuropeptide--receptor relationships. The topology of this network is consequently determined jointly by peptide expression, receptor expression, experimentally characterized ligand--receptor compatibility, and the spatial assumptions of the selected model.

\subsubsection{Functional signal-propagation connectome}
\label{sec:data_functional}

The functional connectivity dataset was obtained directly from the causal signal-propagation atlas published by Randi et al.~\cite{randi2023}.

In the original experiments, identified neurons were stimulated individually using optogenetics while responses in other neurons were simultaneously monitored through calcium imaging. Repeating this perturbation across a large number of neuronal pairs produced a directed map of experimentally observed influence across the nervous system. Only interactions satisfying the false-discovery criterion $q<0.05$ were retained for the present analysis.

An edge therefore represents a statistically supported response in a recorded neuron following stimulation of another identified neuron. The corresponding edge values quantify experimentally measured changes in calcium activity, computed over the 30-s window following stimulation and averaged across trials and animals. These values can contain both positive and negative entries, representing increases or decreases in the downstream response.

The orientation of this matrix differs from the preceding directed datasets. Columns correspond to the stimulated, or pre, neuron, whereas rows correspond to the responding, or post, neuron. Thus, an entry in row $i$ and column $j$ represents a weighted directed connection from the stimulated neuron $j$ to the responding neuron $i$.

The functional network also differs from the other layers in its neuronal coverage. The matrix analyzed here contains 188 neurons represented in the perturbation atlas, determined by the neurons that could be stimulated and recorded within the experimental protocol. Its connectivity describes the interactions supported by this sampling and statistical selection. Unmeasured neuronal pairs and measured pairs without a retained response have distinct experimental interpretations, even though neither contributes an edge to the analyzed graph.

\subsection{Max-plus algebra}\label{sec:maxplus}
 
Max-plus algebra replaces ordinary addition by maximization and ordinary multiplication by addition~\cite{cuninghamegreen1979,gaubert1997}. It is suited to systems in which alternative paths compete and the dominant one is selected, and has been applied to scheduling, discrete-event systems and optimal-path problems~\cite{baccelli1992,gaubert1997}. We summarize here the elements needed to read our results; the reducible spectral theory is given in~\cite{valverde2016} and its use for modeling neural dynamics in a companion paper~\cite{companion}. % TODO keys [60,17,18]
 
\paragraph{Operations and graph interpretation.} We work in the completed max-plus semifield $\cmaxplus$, with carrier $\Rbar=\mathbb{R}\cup\{\pm\infty\}$ and operations
\begin{equation}
a\mplus b=\max(a,b),\qquad a\mtimes b=a+b,
\end{equation}
bottom $\bot=\varepsilon=-\infty$ (neutral for $\mplus$), unit $e=0$ (neutral for $\mtimes$) and top $\top=+\infty$; operations involving $\bot$ and $\top$ follow~\cite{valverde2016}. Matrix operations replace sums by $\mplus$ and products by $\mtimes$; in particular
\begin{equation}
(A\mtimes x)_i=\max_{j}\,(A_{ij}+x_j).
\end{equation}
With $A_{ij}$ the weight of the influence from neuron $j$ to neuron $i$, each neuron receives the largest weighted contribution among its predecessors. Each dataset is embedded as
\begin{equation}
A_{ij}=\begin{cases} w_{j\to i}, & \text{if a retained connection } j\to i \text{ exists},\\ \bot, & \text{otherwise,}\end{cases}
\end{equation}
so that an absent edge is $\bot$, whereas a numerical zero (the unit $e$) or a retained negative response is a finite edge. The entry $(A\mpow{k})_{ij}$ is the largest accumulated weight among walks of length $k$ from $j$ to $i$, and the Kleene star $A^{\mstar}=\msum{k\geq 0} A\mpow{k}$ collects the best walks of any length.

The completed max-plus semifield has an order dual, the completed min-plus semifield $\cminplus$, with $a \nplus b=\min(a,b)$ and $a \ntimes b=a+b$; operators with a dot above belong to it. The two share the carrier $\Rbar$, exchange the roles of $\bot$ and $\top$, and differ only in products involving these extremal elements ($\bot\mtimes\top=\bot$, whereas $\bot\ntimes\top=\top$)~\cite{valverde2016}. Max-plus algebra describes propagation, whereas inverse questions, such as which states are compatible with an observed output, are linear in min-plus algebra~\cite{companion}. The generalized spectral theory used here is formulated for both semifields; we report the max-plus results.
 
\paragraph{Strongly connected components.} The graph $\mathcal{G}(A)$ has one node per neuron and one edge $j\to i$ per finite entry $A_{ij}$. It decomposes into strongly connected components (SCCs), maximal sets of mutually reachable neurons; collapsing each SCC into one node gives the acyclic condensation graph.
 
\paragraph{Eigenvalues, critical cycles and eigenvectors.} The right and left eigenproblems are
\begin{equation}
A\mtimes v=\lambda\mtimes v,\qquad u\tr\mtimes A=\lambda\mtimes u\tr,
\end{equation}
that is, $\max_j(A_{ij}+v_j)=\lambda+v_i$ for every $i$. One propagation step preserves the eigenvector pattern up to a common shift $\lambda$, so only differences between finite coordinates are meaningful. For an irreducible (strongly connected) matrix, the eigenvalue is the maximum cycle mean
\begin{equation}
\lambda(A)=\max_{c\in\mathcal{C}(A)}\frac{w(c)}{\ell(c)},
\end{equation}
where $w(c)$ and $\ell(c)$ are the total weight and the length of cycle $c$. Cycles attaining it are critical; their union is the critical graph, whose SCCs are the critical classes. With $\tilde A=\lambda^{-1}\mtimes A$, the eigenspace $V(A)$ is generated by the columns of $\tilde A^{\mstar}$ indexed by critical nodes, one per critical class: the fundamental eigenvectors~\cite{gaubert1997}.
 
For reducible matrices, every recurrent SCC contributes its own spectral value. We used the constructive procedure of~\cite{valverde2016}, which solves each irreducible block, propagates the solutions through the condensation graph, and returns the complete set of fundamental eigenvectors of the full matrix. In these generalized eigenvectors, $\bot$ marks neurons that cannot be reached from (right) or cannot reach (left) the critical class, and $\top$ marks neurons dominated by a recurrent component with a larger eigenvalue (Section~\ref{sec:maxplus_synaptic}).

\backmatter

\bmhead{Acknowledgements}

This work was funded by Grant PID2024-158251NB-C31 funded by MICIU/AEI/10.13039/501100011033 and by “ERDF/EU” and DAAD project 57616814, SECAI, School of Embedded Composite AI as part of the program Konrad Zuse Schools of Excellence in Artificial Intelligence

\section*{Declarations}

\bmhead{Data availability} All datasets are publicly available: gap-junction and chemical-synapse matrices from WormWiring (corrected July 2020 release)~\cite{wormwiring2020}; monoaminergic network from the \textit{C.~elegans} Connectome Toolbox~\cite{connectometoolbox}; neuropeptidergic network from Ripoll-Sánchez et al.\ (February 2024 update)~\cite{ripollsanchez2023, ripollsanchez2024data}; signal-propagation atlas from Randi et al.~\cite{randi2023}. Processed matrices are included in the code repository. 
 
\bmhead{Code availability} The analysis code is available at \url{https://github.com/emmareyner/Maxplus_eigenstructure_for_neuraldata}. It accepts user-provided connectivity matrices and can be applied to other weighted neuronal networks. 
 
\bmhead{Author contributions} 

\insertcreditsgranularstatement
 
\bmhead{Competing interests} The authors declare no competing interests.

\noindent

\bibliography{references}% common bib file

\bigskip
\begin{appendices}

\section{Quick reading guide to max-plus algebra and its network structures}
\label{app:maxplus_guide}
This appendix provides an intuitive guide to the structures used in the Results. The formal definitions and computational procedures are presented in Methods. Throughout the guide, neurons are represented as nodes and directed interactions as weighted edges. An entry $A_{ij}$ describes an interaction from neuron $j$ to neuron $i$.

\subsection*{What does max-plus analysis measure?}

Max-plus algebra replaces ordinary addition with maximization and ordinary multiplication with addition:

$a\mplus b=\max(a,b),\quad a\mtimes b=a+b.$

For a neuronal network, this means that connection weights are added along a directed route, and alternative routes of the same length are compared by selecting the largest accumulated weight. Recurrent routes can be compared through their mean weight per connection.

For example, suppose that neuron $P$ connects to neuron $Q$ with weight 6, and $Q$ connects back to $P$ with weight 4. The resulting two-edge cycle has total weight 10 and mean weight 5. A self-connection of weight 5 has the same cycle mean. Comparing mean weights accounts for the different lengths of recurrent routes.

The biological interpretation follows the meaning of the original measurements. A large weight may represent extensive anatomical contact, several supported ligand--receptor relationships, or a large measured response. Accordingly, dominance in this analysis refers to the weights and path-composition rule of the network under study.

\subsection*{Strongly connected components: where can recurrence occur?}

A strongly connected component (SCC) is a maximal group of neurons in which every neuron can reach every other neuron through directed paths. The paths can pass through intermediate neurons, so membership in the same SCC does not require a direct connection between every pair.

An SCC containing more than one neuron necessarily contains recurrent routes. A singleton SCC supports recurrence when it has a self-connection. Other singleton components can represent neurons positioned along incoming or outgoing routes, or neurons disconnected from the rest of the network.

Biologically, SCCs identify the neuronal domains within which signals have routes to circulate and return. Their size describes the extent of mutual reachability.

\subsection*{Condensation graph: how are recurrent regions arranged?}

The condensation graph replaces each SCC with a single node. A directed edge between two such nodes indicates that at least one connection links their constituent neurons in that direction.

This representation makes the large-scale organization easier to read. It identifies components that supply incoming routes to a recurrent region, components reached through its outgoing routes, and connections between distinct recurrent regions. The condensation graph is acyclic: a directed cycle between several components would make them mutually reachable and therefore part of the same SCC.

We use \emph{upstream} for neurons or components with paths leading towards a region of interest, and \emph{downstream} for those reachable from it.

\subsection*{Critical cycles and classes: which recurrent circuits dominate?}

Within a recurrent component, a critical cycle is a closed directed route with the greatest mean edge weight. It identifies a particular feedback circuit attaining the maximum average weight per connection.

Several cycles can attain the same maximum. Their union forms the critical graph, and the SCCs of this graph are called critical classes. A critical class can therefore contain several equally dominant cycles connected through shared critical structure.

This distinguishes the full region in which recurrence is possible from the specific circuits that dominate its weighted organization. A large recurrent component can contain a critical class involving only one or a few neurons.

\subsection*{Eigenvalue: what does the dominant cycle weight mean?}

For a strongly connected recurrent component, the max-plus eigenvalue $\lambda$ equals the maximum cycle mean. In the example above, a critical cycle with weights 6 and 4 gives
\[
\lambda=\frac{6+4}{2}=5.
\]

The eigenvalue summarizes the dominant recurrence in the units of the dataset. For anatomical connectivity, it describes the mean anatomical weight along a critical cycle. For molecular signaling networks, it describes the mean number of supported signaling relationships. For the experimental functional network, it describes the mean response weight.

Networks containing several recurrent components can have several spectral values. Their interpretation also depends on the directed connections between those components. Numerical values from datasets with different units are interpreted separately.

\subsection*{Right eigenvectors: which neurons can the critical circuit reach?}

A right eigenvector describes the weighted organization extending downstream from a critical circuit. It relates that circuit to neurons reachable through outgoing paths, including neurons within its own recurrent component and neurons reached beyond that component.

Biologically, this provides a description of the neuronal population towards which the circuit has potential routes of influence. The coordinates incorporate pathway weights relative to the spectral value associated with the circuit.

\subsection*{Left eigenvectors: which neurons can reach the critical circuit?}

A left eigenvector provides the complementary upstream description. It relates a critical circuit to neurons with directed paths leading towards it.

Biologically, this identifies the neuronal population that has potential routes through which to supply signals to the circuit. Neurons in the same recurrent component can belong to both the upstream and downstream domains. Differences between the domains reveal the directional organization of the surrounding network.

\subsection*{How should finite and infinite coordinates be read?}
 
The finite domain of an eigenvector is the set of neurons with finite coordinates. Only differences between finite coordinates are meaningful: adding the same constant to all of them gives the same eigenvector.
 
Infinite coordinates encode structure, not activity. In a right eigenvector, $\bot$ ($=-\infty$) marks neurons that cannot be reached from the critical circuit; in a left eigenvector, neurons that cannot reach it. A coordinate of $\top$ ($=+\infty$) appears when a neuron is reachable from the critical circuit but lies in, or downstream of, another recurrent component whose loops are stronger. In the chemical network, for example, the pharyngeal circuit ($\lambda=15$) reaches the somatic component ($\lambda=37$). Measured relative to the pharyngeal loop, the best accumulated weights in the somatic component grow by 37 per step instead of 15, and the gap increases without bound; the eigenvector records this as $\top$. These values identify which recurrent regime dominates each region of the network.

\subsection*{Reading the structures together}

The structures answer successive questions about the same network. SCCs locate recurrent domains, and the condensation graph shows their directional relationships. Critical cycles identify the dominant weighted feedback circuits within those domains, while eigenvalues summarize their mean weights. Left and right eigenvectors then describe how the circuits connect to upstream and downstream neuronal populations. Together, they link specific recurrent interactions to the organization of the network as a whole.

\end{appendices}

\end{document}